\documentclass[referee]{jfm}

\usepackage{graphicx}
\usepackage{natbib}

\ifCUPmtlplainloaded \else
  \checkfont{eurm10}
  \iffontfound
    \IfFileExists{upmath.sty}
      {\typeout{^^JFound AMS Euler Roman fonts on the system,
                   using the 'upmath' package.^^J}%
       \usepackage{upmath}}
      {\typeout{^^JFound AMS Euler Roman fonts on the system, but you
                   dont seem to have the}%
       \typeout{'upmath' package installed. JFM.cls can take advantage
                 of these fonts,^^Jif you use 'upmath' package.^^J}%
      }
  \else
  \fi
\fi

\ifCUPmtlplainloaded \else
  \checkfont{msam10}
  \iffontfound
    \IfFileExists{amssymb.sty}
      {\typeout{^^JFound AMS Symbol fonts on the system, using the
                'amssymb' package.^^J}%
       \usepackage{amssymb}%

      }{}
  \fi
\fi

\ifCUPmtlplainloaded \else
  \IfFileExists{amsbsy.sty}
    {\typeout{^^JFound the 'amsbsy' package on the system, using it.^^J}%
     \usepackage{amsbsy}}
    {\providecommand\boldsymbol[1]{\mbox{\boldmath $##1$}}}
\fi

\newcommand\Real{\mbox{Re}} 
\newcommand\Imag{\mbox{Im}} 
\newcommand\Rey{\mbox{\textit{Re}}}  

\newsavebox{\astrutbox}
\sbox{\astrutbox}{\rule[-5pt]{0pt}{20pt}}

\newcommand\p{\ensuremath{\partial}}
\newcommand\tti{\ensuremath{\rightarrow\infty}}

\newcommand\ttz{\ensuremath{\rightarrow 0}}
\newcommand\tto{\ensuremath{\rightarrow 1}}

\newcommand\eg{e.g.\ }
\newcommand\ie{i.e.\ }

\title[Stability of a laminar mixing layer]{The critical Reynolds
  number of a laminar mixing layer}

\author[P. Bhattacharya, M. P. Manoharan, R. Govindarajan and
  R. Narasimha]%
{P\ls I\ls N\ls A\ls K\ls I\ns B\ls H\ls A\ls T\ls T\ls A\ls C\ls H\ls 
  A\ls R\ls Y\ls A,\ns M.\ns P.\ns M\ls A\ls N\ls O\ls H\ls A\ls
  R\ls A\ls N\thanks{Present address: D4/218, BHEL
  Township, Tiruchirapalli 620014, India.},\break R\ls A\ls M\ls A\ns
  G\ls O\ls V\ls I\ls N\ls D\ls A\ls R\ls A\ls J\ls A\ls N \and R.\ns
  N\ls A\ls R\ls A\ls S\ls I\ls M\ls H\ls A}

\affiliation{Engineering Mechanics Unit, Jawaharlal Nehru Centre
  for Advanced Scientific Research, Jakkur, Bangalore 560064,
  Karnataka, INDIA}

\pubyear{1996}
\volume{538}
\pagerange{119--126}
\date{?? and in revised form ??}

\begin{document}

\maketitle

\begin{abstract}
It has hitherto been widely considered that a mixing layer is unstable
at all Reynolds numbers. However this is untenable from energy
considerations, which demand that there must exist a non-zero Reynolds
number below which disturbances cannot extract energy from the mean
flow. It is shown here that a linear stability analysis of similarity
solutions of the plane mixing layer, including the effects of flow
non-parallelism, using the minimal composite theory and the properties
of adjoints following \cite{Govin05}, resolves the issue by yielding
non-zero critical Reynolds numbers for coflowing streams of any
velocity ratio. The critical Reynolds number so found, based on the
vorticity thickness and the velocity differential as scales, varies in
the narrow range of $59$ to $55$ as the velocity ratio goes from zero
to unity.
\end{abstract}

\section{Introduction}\label{sec:intro}
The motivation for the present work arises from an analysis of the
stability of a mixing layer due to \cite{Betchov63}. This analysis,
based on the Orr--Sommerfeld (OS) equation, showed that the neutral
curve in the wave-number ($\alpha$) -- Reynolds number (\Rey) plane
approaches the origin $\alpha=0$ as \Rey\ goes to $0$, leading the
authors to conclude ``No minimum [\ie critical] Reynolds number is
found'' for the flow. Mixing layers have been studied extensively
since, and recent texts, \eg \citet[pp.~81--294]{Huerre98},
\citet[p.~90]{Crimin03} and \citet[p.~199--201]{Drazin04}, still
report the critical Reynolds number as zero. This is intriguing, for
several early studies of stability (\eg \citet[pp.~178--183]{Pran35},
\citet[pp.~31--32, 59+]{Lin55}, going back to the two-dimensional
analysis of \citet[pp.~43--71]{Loren07}), show that (under certain
reasonable conditions) a two-dimensional incompressible viscous flow
must be stable at sufficiently low Reynolds numbers. Astonishingly,
however, there is no analysis in the extensive literature on
hydrodynamic stability that yields a non-zero value for this critical
Reynolds number. The mixing layer being a very basic flow type, the
absence of any definitive commentary on the above issue has led other
studies of flows modelled on the mixing layer \cite*[\eg][]{Solom93,
  Ostek99} to base their analyses on the assumption that the critical
Reynolds number is zero.

Now all analyses of mixing layer stability, from \cite{Esch57} to
\cite{Balsa87}, start with the OS equation, which is
valid only for strictly parallel flow. However, the width of a laminar
mixing layer scales as $x^{1/2}$ (where $x$ is the streamwise distance),
so the rate of change of thickness (and hence also any parameter
measuring the degree of flow non-parallelism) becomes infinite in the
limit of $x$ (equivalently \Rey) going to $0$. In other words,
existing studies are based on the assumption of no non-parallelism in
a situation where any measure of non-parallelism would be
infinite. Thus flow non-parallelism may be expected to play a crucial
role in determining the stability characteristics in the limit. Though
\cite{Crimin94} note that ``with viscous effects, the basic flow
should be treated as non-parallel'', and other texts
\cite[\eg][p.197]{Drazin04} emphasize that ``at small values of
Reynolds number the parallel-flow assumption is of questionable
validity'', no investigation accounts for the non-parallelism. We show
here that a consistent non-parallel flow theory yields finite,
non-zero critical Reynolds numbers based on vorticity thickness and
velocity differential in the range $55$ to $59$ depending on the
velocity ratio.

In a departure from earlier work, the similarity solution of the
laminar mixing layer is used as the base flow everywhere in the
present analysis. The earlier studies, beginning with ones that
assumed zero viscosity, have moved from the discontinuous profile due
to Helmholtz to the hyperbolic tangent profile considered by
\cite{Betchov63}. Examples of other approximations appear in
\cite{Esch57}, and all later studies to our knowledge use one of
these. The similarity profile is a better approximation of reality
than any of the above, and its use through what we have called minimal
composite theory is both appropriate and convenient in the present
approach.

The rest of the paper is arranged as follows. In~\S\,\ref{sec:similar}
the similarity solution of the mean flow profile is presented and some
remarks follow. The stability problem is posed and the method of
solution briefly outlined in~\S\,\ref{sec:stabil}, essentially
following the approach used for boundary layers by
\cite{Govin97,Govin05}. In~\S\,\ref{sec:results}, the results of the
stability analysis are presented and compared with earlier work.

\section{Similarity solution of the basic flow}\label{sec:similar}
A plane incompressible mixing layer that develops in the positive
$x$-direction is considered. The two free-streams flow with velocities
$U_\infty$ and $\lambda U_\infty$ (with $\lambda < 1$) before
coming into contact with each other at the origin ($x=0$). Both
streams are semi-infinite in their lateral ($y$) extent. We omit the
case of counterflowing streams ($\lambda < 0$) from the present
discussion, since they are not relevant to the non-parallel flow
theories. Since no other external length scale is present in the
problem as formulated, after a sufficient distance from the origin
similarity may be taken to be valid.

Henceforth all variables subscripted by {\em d} are dimensional
quantities. The streamwise coordinate and the coordinate in the
direction normal to the flow are $x_d$ and $y_d$ respectively, and
$\nu$ is the kinematic viscosity. The similarity analysis here differs
from \citet[][pp.~175--176]{Schlich04} only in the definition of the
length scale $l_d$ in the direction normal to the flow,
\begin{equation}
l_d(x_d)\equiv\displaystyle\sqrt{\nu x_d \over U_\infty}.
\label{eqn:length}
\end{equation}
The streamfunction is defined as 
\[
 \mathit{\Phi}_d (x_d,y_d) = U_\infty l_d(x_d) \mathit{\Phi}(y),
\]
$y=y_d/l_d$ being the similarity coordinate. The momentum equation
in terms of the non-dimensional streamfunction $\mathit{\Phi}$ becomes
\begin{equation}
 \mathit{\Phi}''' + {1\over2}\mathit{\Phi}\mathit{\Phi}'' = 0,
\label{eqn:simileq}
\end{equation}
where the primes denote derivatives taken with respect to $y$. The
boundary conditions to be satisfied are
\begin{equation}
\left. \begin{array}{ll}
\mathit{\Phi}'(y=+\infty) = 1,\\[8pt]
\mathit{\Phi}'(y=-\infty) = \lambda \qquad \mbox{ and}\\[8pt]
\left.(y\mathit{\Phi}'-\mathit{\Phi})\right]_{y=\infty} +
\lambda\left.(y\mathit{\Phi}'-\mathit{\Phi})\right]_{y=-\infty}=0. 
 \end{array}\right\}
\label{eqn:similbc}
\end{equation}
The third boundary condition has been a subject of some
controversy. As formulated above, it is derived from matching the
pressure across the mixing region, following \cite{Ting59}. The zero
net transverse force condition suggested by \cite{Karman21} results in
an identical third boundary condition. \cite{Acriv72} pointed out
an inconsistency in the above formulation and showed that this
condition is still incomplete, and further that it cannot be resolved
within the context of classical boundary layer theory. Using a
different approach to formulate asymptotic expansions to solutions of
the Navier--Stokes equations, \cite{Alston92} have however argued that
this condition remains the best option compared to the other
alternatives used in the literature.

\begin{figure}
  \centerline{\includegraphics[angle=0,origin=c,scale=0.40]{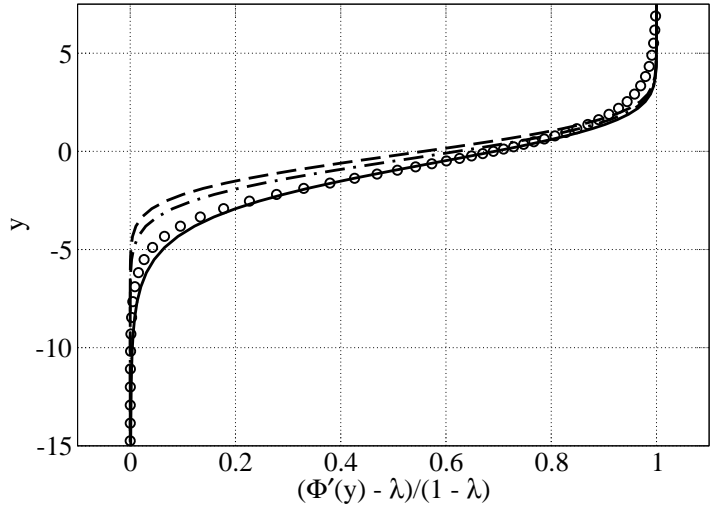}}
  \caption{Streamwise velocity profiles from the solution
  of~(\ref{eqn:simileq}) for different velocity ratios. Solid line,
  similarity profile for $\lambda=0$; circles, ${1\over
  2}(1+\tanh\{2(y+1)/5\})$; dash-dotted line, similarity profile for
  $\lambda=0.5$; short dashes, similarity profile for
  $\lambda=0.75$.}\label{fig:velpro}
\end{figure}

The velocity profiles obtained for different values of $\lambda$ are
shown in figure~\ref{fig:velpro}. Apart from the displacement of the
dividing streamline in the negative $y$ direction, these solutions are
identical to those obtained first by \cite{Lock51}. The result for the
{\em half jet}, \ie for $\lambda =0$, was validated against that given
in \citet[pp.~175--176]{Schlich04}. A suitably transformed hyperbolic
tangent function is superimposed on the similarity profile for
$\lambda=0$ for comparison.

Note that for every flow configuration with a $\lambda$ in the range
$0$ to $1$, there corresponds an identical but vertically flipped flow
with velocity ratio $1/\lambda$. So it is expected that the similarity
solution becomes increasingly symmetric about the $x$ axis as
$\lambda$ approaches $1$. In the limit $\lambda\tto$,
redefining the non-dimensional streamfunction as
\[
\mathit{\Phi}(y)\equiv(1-\lambda)f(y) + y,
\]
the similarity equation~(\ref{eqn:simileq}) becomes
\begin{equation}
f''' + {1 \over 2}[y + (1-\lambda)f]f'' = 0,
\label{eqn:newsimil}
\end{equation}
with appropriately transformed boundary conditions. Writing
($1-\lambda$) as a small parameter $\epsilon$ and assuming a solution
in the form of an asymptotic series in powers of $\epsilon$,
\[
f = f_0 + \epsilon f_1 + \epsilon^2 f_2 + \mbox{ (higher order terms)},
\] 
we get the streamwise velocity correct to second order in $\epsilon$
as
\[
U_{d0}(x_d,y_d) = U_\infty[1+ \epsilon f_0'(y)] =
U_\infty\biggl\{1 - {\epsilon \over 2}\biggl(1 -
   \mbox{erf}(y/2)\biggr)\biggr\} + O(\epsilon^2).
\]
While such analytically expressible profiles are ubiquitous in the
literature, our purpose is to demonstrate that the similarity solution
smoothly merges into the error-function profile for $\lambda$ close to
$1$. It is therefore not expected (as we shall confirm shortly) that
the stability characteristics change very much between these
choices. This puts into perspective one of the nuances of our
analysis that is different from existing ones, for it must be
emphasized that the marked deviation observed in the final result is
not to be attributed to the differences in the assumed velocity
profiles but to the non-parallel flow analysis.

\section{The non-parallel stability problem}\label{sec:stabil}
A brief outline of the minimal composite theory is described in what
follows. A review of the method can be found in \cite{Naras00}.

Each flow quantity, for example the streamfunction, is expressed as
the sum of a mean $\mathit{\Phi}_d$ and a perturbation
$\skew4\hat\phi_d$, where 
\begin{eqnarray}
\mathit{\Phi}_d(x_d,y_d)&=&U_{\infty}l_d(x_d)\mathit{\Phi}(y)
  \qquad\mbox{ and} \\ 
\skew4\hat\phi_d(x_d,y_d)&=&U_{\infty}l_d(x_d)\phi(x,y)\exp
  \biggl[\mathrm{i}\biggl(\int\alpha(x)\mathrm{d}x-\omega t\biggr)
  \biggr], 
\end{eqnarray}
with the (complex) phase speed of the disturbance
$c=\omega/\alpha$. Inserting these into the Navier--Stokes equations
for two-dimensional incompressible flow written in terms of the
streamfunction, and retaining all terms nominally upto $O(\Rey^{-1})$,
the non-parallel stability equation can be written as
\begin{equation}
\label{eqn:fullneq}
\mathcal{N}\{\phi\} = 0, \qquad\mbox{ with boundary conditions} \quad
\phi,\mathrm{D}\phi\ttz\quad\mbox{ at } y=\pm\infty,
\end{equation}
where, for the mixing layer,
\begin{eqnarray*}
\mathcal{N} &\equiv& (\mathit{\Phi}'-c) (\mathrm{D}^2 - \alpha^2) -
\mathit{\Phi}''' + \frac{\mathrm{i}}{\alpha \Rey}\biggl\{\mathrm{D}^4 -
2\alpha^2\mathrm{D}^2 + \alpha^4 + p \mathit{\Phi} \mathrm{D}^3 \biggr.\\
&& + p (\mathit{\Phi}'\mathrm{D}^2+\mathit{\Phi}''\mathrm{D}) -
\alpha^2[2py(\mathit{\Phi}'-c) + p\mathit{\Phi}]\mathrm{D}
-p\alpha^2c + p\mathit{\Phi}'''\\
&& \biggl. + (3\mathit{\Phi}'-c)\Rey\alpha\alpha' + [\mathit{\Phi}'''
  + \alpha^2 (3\mathit{\Phi}'-2c) -
  \mathit{\Phi}'\mathrm{D}^2]\Rey\frac{\p}{\p x}\biggr\},
\end{eqnarray*}
$\mathrm{D}$ is the derivative with respect to $y$ and
$p\equiv\Rey(\mathrm{d}l_d/\mathrm{d}x_d)=1/2$. The Reynolds number is
based on $U_\infty$ and $l_d$. Note that in the minimal composite
theory the operator $\mathcal{N}$ is flow-specific, and so is
different \eg from that for the boundary layer (see below). Following
\cite{Govin97} the non-parallel operator $\mathcal{N}$ is expressed as
the sum of an operator $\mathcal{M}$ that contains all the lowest
order terms and an operator $\mathcal{H}$ comprising the higher order
terms,
\begin{eqnarray*}
\mathcal{M} &\equiv& (\mathit{\Phi}'-c) (\mathrm{D}^2 - \alpha^2) -
\mathit{\Phi}''' + \frac{\mathrm{i}}{\alpha \Rey} \mathrm{D}^4,\\ 
\mathcal{H} &\equiv& \frac{1}{\Rey}\biggl\{p \mathit{\Phi}
\mathrm{D}^3 -2\alpha^2\mathrm{D}^2 + \alpha^4 +
p (\mathit{\Phi}'\mathrm{D}^2+\mathit{\Phi}''\mathrm{D}) -
\alpha^2[2py(\mathit{\Phi}'-c) + p\mathit{\Phi}]\mathrm{D} \\
&& -p\alpha^2c + p\mathit{\Phi}''' \biggr\} +
(3\mathit{\Phi}'-c)\alpha\alpha' + \mathcal{S}\frac{\p}{\p x}, \\
\mbox{where }\;\qquad\mathcal{S} &\equiv& [\mathit{\Phi}''' +
  \alpha^2(3\mathit{\Phi}'-2c) -  \mathit{\Phi}'\mathrm{D}^2].
\end{eqnarray*}
The relative order of magnitude of each individual term varies with
$y$ because of the presence of the critical layer at
$\mathit{\Phi}'=c$. While constructing the minimal composite equation
that yields results correct upto $O(\Rey^{-1})$, the order of
magnitude of any term within the $y$-domain is considered. The final
equation consists of all terms that are at least $O(\Rey^{-1})$
{\em somewhere}, and rejects all that are $o(\Rey^{-1})$ {\em
  everywhere} in the domain. It must be emphasized that the above
equations are different from those in \cite{Govin05},
which were formulated for a boundary layer. For instance, in the
construction of the operator $\mathcal{M}$ for the present case
clearly no wall layer considerations are needed. Furthermore, the term
$p\mathit{\Phi}\mathrm{D}^3$, being of higher order everywhere in the
mixing layer, is in the operator $\mathcal{H}$, whereas in the
boundary layer it is part of $\mathcal{M}$. The solution procedure for
estimating the growth of the disturbance, though, remains the same,
and is given in that paper in detail. For reference, we recapitulate
some of its essential points below.

It is instructive to note that though the non-parallel operator has
partial derivatives in both $x$ and $y$, the lowest order terms
(comprising $\mathcal{M}$) contain derivatives only in $y$. Taking
this as a cue, the total solution $\phi$ is expressed in two parts,
\begin{equation}
\label{eqn:fullphi}
\phi(x,y) = A(x)\phi_m(x,y) + \epsilon \phi_h(x,y).
\end{equation}
Here $A(x)$ is the amplitude function that captures the streamwise
variation of the lowest order solution $\phi_m$, which satisfies the
equation
\begin{equation}
\mathcal{M}\{\phi_m\}=0.
\label{eqn:minimal}
\end{equation}
We consider the downstream growth of disturbances at a constant value
of the similarity variable $y$. For future reference, we also define a
complex {\em effective} wavenumber associated with the non-dimensional
streamwise disturbance velocity $\skew2\hat u$ as
\begin{equation}
\alpha_{\rm{eff}} = \alpha -\mathrm{i}\biggl({1 \over A} {\mathrm{d}A
  \over \mathrm{d}x} + {1 \over \p \phi_m /\p y} {\p (\p \phi_m / \p
  y) \over \p x}\biggr),
\label{eqn:effalpha}
\end{equation}
such that the growth rate $g$ of $\skew2\hat u$ is given by 
\begin{equation}
g \equiv -\Imag(\alpha_{\rm{eff}}) = -\alpha_i + \Real\biggl({1 \over
  A} {\mathrm{d}A \over \mathrm{d}x} + {1 \over \p \phi_m /\p y} {\p
  (\p \phi_m / \p y) \over \p x}\biggr),
\label{eqn:growth}
\end{equation}
where $\alpha_i$ is the imaginary part of the full complex wavenumber
$\alpha$. We find that, in general, a disturbance may amplify at one
$y$ and decay at another. Moreover, one disturbance quantity could be
amplifying while others decay. The derivative of $\phi_m$ with respect
to $x$ is obtained by solving~(\ref{eqn:minimal}) for a nearby $\Rey$
and noting that $\p/\p x = p \p/\p \Rey$. Substituting
from~(\ref{eqn:fullphi}) into~(\ref{eqn:fullneq}), and noting that
$\mathcal{N} = \mathcal{M} + \mathcal{H}$, we obtain the amplitude
evolution equation
\begin{equation}
A\mathcal{H}\phi_m + {\mathrm{d}A \over \mathrm{d}x}\mathcal{S}\phi_m
= -\epsilon\mathcal{M}\phi_h,
\label{eqn:ampevol}
\end{equation}
the truncated terms here being $O(\Rey^{-2/3})$ compared to the largest
of the retained terms. Further the expression for the adjoint of the
operator $\mathcal{M}$ is found to be
\[
\mathcal{\overline{M}}=(\mathit{\Phi}'-c^*)(\mathrm{D}^2-\alpha^{*2}) +
2\mathit{\Phi}''\mathrm{D} - {\mathrm{i}\over\alpha^*\Rey}\mathrm{D}^4
+ O(\Rey^{-2/3}),
\]
where a quantity superscripted with an asterisk represents its complex
 conjugate. Using the property of adjoints \cite[cf.][]{Govin05}, the
 contribution to the growth $g$ due to change in the amplitude
 function, \ie the quantity $(1/A)\mathrm{d}A/\mathrm{d}x$, is
 calculated without the need to specifically compute the higher order
 solution $\phi_h$. Equation~(\ref{eqn:ampevol}) therefore reduces to
\[
A\int_{-\infty}^{\infty}\chi^*\mathcal{H}\{\phi_m\}\mathrm{d}y +
  {\mathrm{d}A\over\mathrm{d}x} \int_{-\infty}^{\infty}
  \chi^*\mathcal{S}\{\phi_m\} \mathrm{d}y = \mbox{ (higher order
  terms)},
\]
where $\chi$ is the solution of the adjoint problem,
\[
\mathcal{\overline{M}}\{\chi\}=0.
\]
Now $A\sim O(1)$ and $\mathrm{d}A/\mathrm{d}x$ is $O(\Rey^{-1})$, so
the error in the estimation of the growth rate will be
$o(\Rey^{-1})$. But as we integrate over a large streamwise distance
of $O(\Rey)$, the error in the amplitude is expected to be
$o(1)$. Here, a solution of the partial differential equation
(\ref{eqn:fullneq}) to the desired order of accuracy has been obtained
by solving a parametric ordinary differential equation and using the
property of adjoints. An obvious advantage is that the complexity of
the problem is significantly reduced when compared to the exercise of
solving it as a full partial differential equation.

\section{Results}\label{sec:results}

\subsection{Accuracy of results}\label{sec:accuracy}
Both the parallel and non-parallel stability analyses involve solving
an eigenvalue problem. The conditions at infinity are posed at
distances sufficiently far away to ensure that results are independent
of the size of the domain. For example, on the curve of marginal
stability, the Reynolds number for a given $\alpha$ is required to be 
identical up to the fifth decimal place before any further increase in
domain size is considered unnecessary. We notice that at large $y$,
$\mathit{\Phi}'''$ vanishes and the higher derivatives of the
eigenfunction with respect to $y$ (appearing in the viscous term)
become smaller. Therefore, as an estimate, the eigenfunction decays as
$\exp(\pm\alpha y)$, \ie its $e$-folding rate depends on the
wavenumber of the disturbance. The $y$-domain thus needs to be larger
for smaller wavenumbers. Further, since the eigenfunctions are
discretized as eigenvectors, a sufficient number of grid-points must
be contained in the domain so that the results are independent of
resolution.

Now, an increase in the number of grid-points directly increases the
size of the matrices involved in the eigenvalue problem and this
affects the computational effort adversely. For non-dimensional
wavenumbers larger than $0.2$, we impose outer boundary conditions at
$y=\pm20$ and use $81$ grid-points with a $\sinh$ grid-stretching. For
the smallest wavenumbers considered ($\alpha\simeq0.02$) the conditions
at infinity are imposed at $y=\pm300$ with $150$
grid-points. Intermediate values were taken to expedite the solution 
process whenever this did not affect the accuracy above a tolerance
level of five significant digits. There is thus a lower cut-off for
the real part of the wavenumber in our results, beyond which obtaining
numerically accurate results becomes prohibitively expensive
computationally. The neutral boundary is defined where the imaginary
part of the eigenvalue $\omega$ and the growth rate $g$ are smaller
than $1\times10^{-8}$ and $1\times10^{-6}$ in magnitude
respectively. In calculating the streamwise derivatives, nearby
stations are taken corresponding to values of Reynolds number $\Rey$
apart by $1\%$ of the value at either of these points. Independence of
the curve of marginal stability to small deviations in the numerical
value of this parameter has also been confirmed.

\subsection{Parallel vs. non-parallel analysis}\label{sec:nonvspar}
As mentioned before, the non-parallel approach is expected to deviate
significantly from the parallel approach in the region where the
Reynolds number based on streamwise distance is small.
\begin{figure}
  \centerline{\includegraphics[angle=-90,origin=c,scale=0.40]{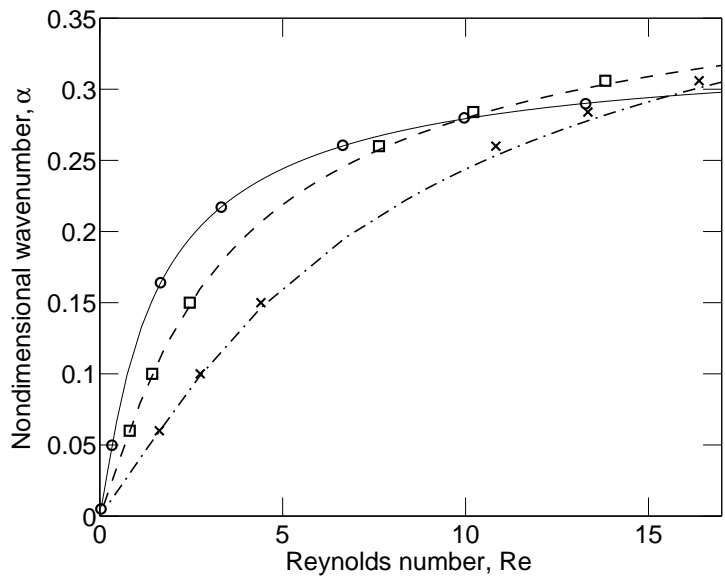}}
  \caption{Results of OS (parallel) analysis: circles, Betchov \&
    Szewczyk (1963); solid line, present analysis on $\tanh$ profile;
    squares, transformed $\tanh$ profile with $\lambda = 0$; short
    dashes, similarity profile with $\lambda = 0$; crosses,
    transformed $\tanh$ profile with $\lambda = 0.5$; dash-dot line,
    similarity profile with $\lambda = 0.5$.} 
  \label{fig:betc}
\end{figure}

First, as validation of the OS solver used, we compare the
results of the present analysis on the hyperbolic tangent profile (the
case of counterflowing streams with $\lambda=-1$) with those of
\cite{Betchov63} in figure~\ref{fig:betc}. Secondly, we establish that
the result showing the flow to be unstable at $\Rey = 0$ is not
specific to a counterflow situation, rather it is inherent in the
parallel flow assumption. Curves of marginal stability from a parallel
analysis on the similarity profiles for the velocity ratios $0$ and
$0.5$ are also plotted in figure~\ref{fig:betc}. These are compared
with those from a parallel analysis on a suitably transformed
hyperbolic tangent profile.

Note that while the similarity solution has a well-defined length
scale~(\ref{eqn:length}) associated with it, the hyperbolic tangent
profile may be arbitrarily scaled. For the results shown in
figure~\ref{fig:betc}, the hyperbolic tangent function used is so
transformed that its vorticity thickness~(\ref{eqn:vort}) is
identical to that of the similarity solution for the corresponding
velocity ratio. We note that all the curves approach the origin,
irrespective of whether the hyperbolic tangent or the similarity
profile is used.
\begin{figure}
  \centerline{\includegraphics[angle=-90,origin=c,scale=0.40]{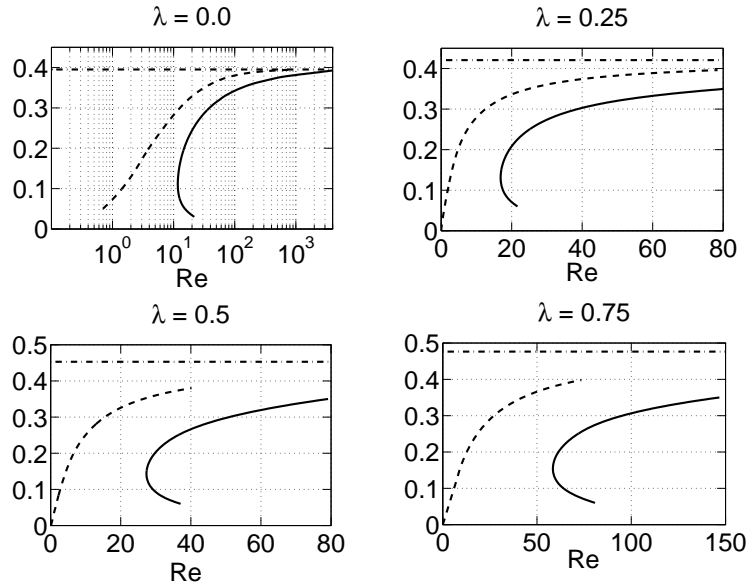}}
  \caption{Comparison of the parallel and non-parallel
    analyses. The Reynolds number is plotted on the horizontal
    axis. The vertical axis corresponds to the real parts of the 
    wavenumber $\alpha$ and the {\em effective} wavenumber 
    $\alpha_{\rm{eff}}$ for the parallel and non-parallel analyses
    respectively. Short dashes, OS; solid line,
    non-parallel; dash-dot line, Rayleigh.}\label{fig:nonpar}
\end{figure}

Next, the results of the parallel and non-parallel analyses are
 compared in figure~\ref{fig:nonpar} for the four velocity
 ratios $0.0, 0.25, 0.5$ and $0.75$. The results correspond to the
 respective similarity profiles. The relevant quantity on the
 vertical axis for the results of the parallel analysis is the
 physical wavenumber $\alpha$, whereas for the non-parallel analysis
 it is the real part of the {\em effective} wavenumber
 $\alpha_{\rm{eff}}$ defined in~(\ref{eqn:effalpha}). 

Apart from the terms already present in the OS equation,
the non-parallel operator $\mathcal{N}$ includes further terms that
essentially provide a correction at low \Rey. It is expected that the
difference between the parallel and non-parallel approaches diminishes
in the limit \Rey\ \tti. Figure~\ref{fig:nonpar} shows that the curves
of marginal stability, from both parallel and non-parallel analyses,
do indeed approach the neutrally stable mode of the solution to
Rayleigh's equation for each of the velocity ratios considered
above. Note that the curves for $\lambda=0$ have been plotted with
\Rey\ on a log scale to show that the results are indistinguishable as
\Rey\ \tti; but differences are noticeable even at $\Rey=10^3$!

From~\S\,\ref{sec:stabil} it is evident that the curve of marginal
stability and hence the critical Reynolds number can vary depending on
the $y$ value at which the growth rate $g$ is determined. The
monitoring location makes a huge difference to the stability result in
the case of boundary layers \cite[][]{Govin97}. But for the present
flow the contribution from the last term in $g$ is smooth in $y$. The
critical $\Rey^*$~(\ref{eqn:resc}) was observed to vary within a
range that was about $10\%$ of the value of the critical $\Rey^*$ at
$y=0$ while traversing from $y=-1.0$ to $y=0.5$. We do not expect any
widely different behaviour outside this range as the variation in the
velocity profile becomes negligible. It is also interesting to note
that the minimum of the critical Reynolds numbers for any given
velocity ratio seems to occur close to the $y$-location where the
streamwise velocity profile has the maximum slope (see
figure~\ref{fig:velpro}). With $y$ rescaled with respect to the
vorticity thickness, for the four velocity ratios considered, the
location of minimum critical $\Rey^*$ deviates most from the location
of maximum slope for $\lambda=0.25$ (by $0.019$ units) and the least
for $\lambda=0$ (by $0.001$ units). All the results from the
non-parallel analysis presented in this paper correspond to a
monitoring location fixed at $y=0$.

\subsection{Critical Reynolds numbers}\label{sec:critre}
\begin{figure}
  \centerline{\includegraphics[angle=-90,origin=c,scale=0.40]{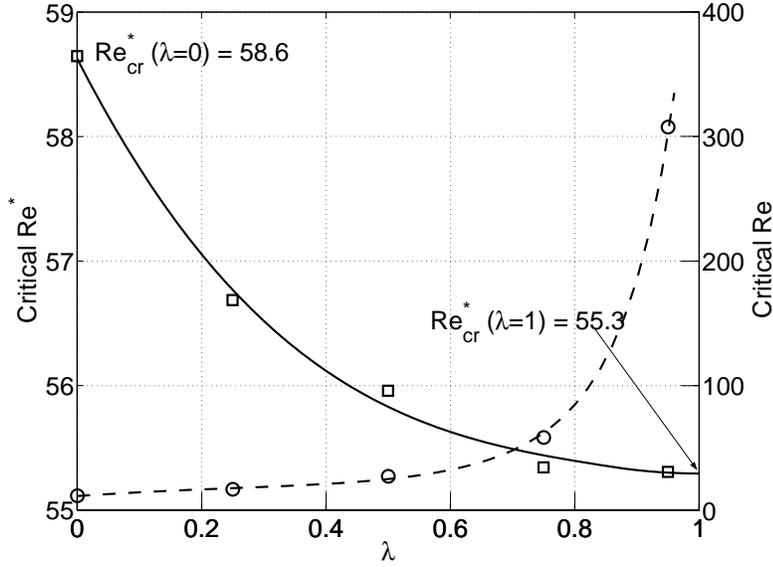}}
  \caption{Variation of the critical Reynolds number with velocity
  ratio. Circles, $\Rey=U_\infty l_d/\nu$; squares,
  $\Rey^*$~(\ref{eqn:resc}). (Dashed and solid curves are fits through 
  the data points.)}\label{fig:recr}
\end{figure}

The variation of critical Reynolds number \Rey\ with velocity ratio is
shown in figure~\ref{fig:recr}. A more appropriate Reynolds number in
the present flow would be one defined in terms of the velocity
difference and the vorticity thickness,
\begin{equation}
\Rey^* \equiv {(1-\lambda)U_\infty \delta_d\over \nu} = {(1-\lambda)^2
  \over\mathit{\Phi''_{max}}}\Rey
\label{eqn:resc}
\end{equation}
where $(1-\lambda)U_\infty$ is the velocity difference, and
\begin{equation}
\delta_\omega(x) \equiv l_d(x) {(1-\lambda) \over \mathit{\Phi''_{max}}}
\label{eqn:vort}
\end{equation}
is the vorticity thickness determined by the maximum slope of the
velocity profile, $\mathit{\Phi''_{max}}$.

The variation of critical $\Rey^*$ with velocity ratio is shown in
figure~\ref{fig:recr}. It varies monotonically with $\lambda$ and
approaches a finite limiting value of $55.3$ as $\lambda$ goes to
$1$. Also, the variation of critical $\Rey^*$ with $\lambda$ is found
to be much less than the variation in critical \Rey.

\section{Conclusions}\label{sec:discnconcl}
The main result of the present work is the demonstration of the
existence of a non-zero critical Reynolds number for a laminar plane
incompressible mixing layer. Choosing the vorticity thickness and
velocity differential as length and velocity scales, this critical
Reynolds number varies in the narrow range from $58.6$ at $\lambda =
0$ to $55.3$ at $\lambda = 1$. The present result puts into
perspective the prevalent understanding of linear stability of the
mixing layer. It also underlines the relevance of a non-parallel
analysis vis-a-vis a parallel one in regard to this flow and other
open flows exhibiting a high degree of non-parallelism. This is a
striking example of a flow where the use of non-parallel theory is
essential to avoid drawing incorrect physical conclusions. Though
parallel flow theory has given revealing insights to instability
mechanisms for over a century, there are regimes of flow where this
theory is qualitatively wrong.

The finding of a critical Reynolds number which is not too low has an
appealing consequence. The laminar similarity flow analysed here is 
physically realizable only at some distance downstream of the splitter
plate, and the results may therefore be verified experimentally or by
direct numerical simulations. This is not possible for the earlier
parallel flow results. We note that since the flow is convectively 
stable below $\Rey^*=55$, the question of absolute instability below 
this Reynolds number does not arise. While it is outside the thrust
of the present work, it is relevant to mention that for inviscid flow, 
from a parallel analysis, it is well established \cite[][]{Huerre85}
that the instability is convective for any mixing layer formed by
coflowing streams.

We also claim that for the purpose of stability analysis, the
dependence of results on the exact velocity profile is weak, the {\em
form} suffices to obtain results to a consistently good accuracy. For
all the non-parallel, parallel and inviscid analyses carried out on
the similarity profiles, the results differ from corresponding
analyses on the (suitably rescaled) hyperbolic tangent profiles by no
more than $5\%$. This is in contrast to wall-bounded shear flows
where the stability results are very sensitive to the mean-flow
velocity profile.

\begin{acknowledgments}
The authors wish to thank the Defence Research and Development
Organisation (DRDO), India for supporting this work.
\end{acknowledgments}

\bibliographystyle{jfm}
\bibliography{jfm_2005}

\end{document}